\documentclass[11pt]{article}

\usepackage[latin1]{inputenc}
\usepackage[T1]{fontenc}
\usepackage[english]{babel}
\usepackage{amsfonts}
\usepackage{amssymb}
\usepackage{amsmath}
\usepackage{amsthm}
\usepackage{graphicx}

\def\be{\begin{equation}}
\def\ee{\end{equation}}
\def\bea{\begin{eqnarray}}
\def\eea{\end{eqnarray}}

\begin{document}

\title{Mean-field cooperativity in chemical kinetics}

\author{\ Aldo Di Biasio\footnote{Dipartimento di Fisica,
   Universit\`a di Parma}
\ Adriano Barra \footnote{Dipartimento di Fisica,
   Sapienza Universit\`a di Roma}
   \ Raffaella Burioni \footnotemark[1]
 \  Elena Agliari \footnotemark[1]
 }

\maketitle  

\begin{abstract}
We consider cooperative reactions and we study the effects of the interaction strength among the system components on the reaction rate, hence realizing a connection between microscopic and macroscopic observables. Our approach is based on statistical mechanics models and it is developed analytically via mean-field techniques. First of all, we show that, when the coupling strength is set positive, the model is able to consistently recover all the various cooperative measures previously introduced, hence obtaining a single unifying framework. Furthermore, we introduce a criterion to discriminate between weak and strong cooperativity, based on a measure of ``susceptibility''. We also properly extend the model in order to account for multiple attachments phenomena: this is realized by incorporating within the model $p$-body interactions, whose non-trivial cooperative capability is investigated too.
\end{abstract}


\section{Introduction}

The phenomenon of cooperativity is widespread in biological and chemical sciences and it has been the focus of many theoretical and experimental investigations \cite{thill}. Cooperativity is a typical "emergent" property, that directly links the description of a system at the single molecular, elementary level, with the macroscopic properties in complex macromolecules, cells and organisms. It is often a compelling task to exploit cooperative effects, such as amplification effects and high sensitivity to external parameters \cite{amit}\cite{bialek}.

A common feature of cooperative systems is the interaction among "active sites": This interaction can result, for example, in an increasing affinity for binding of substrate as more sites are occupied, referred to as positive cooperativity for binding; Conversely, if successive binding of substrate to active sites reduces the binding affinity and inhibits the occupation capability of another site, we speak of negative cooperativity for binding.

Here we focus on the effects of cooperation on reaction rate curves: when measuring the progress of a reaction as a function of the concentration of substrate (e.g. the saturation of hemoglobin by oxygen)  \cite{cook,enzimi} one finds that cooperation is typically associated to sigmoidal curves, in contrast with the hyperbolic Michaelis-Menten law \cite{karlin}, holding when cooperativity is absent. Actually, the plethora of different phenomena where cooperativity effects have been observed  in reaction rates is so spread that  many different definitions have been suggested in the past, most of them being based on the measure of some kind of deviations from the classical Michaelis-Menten reaction kinetics \cite{karlin}. While non-cooperativity is a well defined behavior in complex multi-sites binding,  all the definitions of cooperativity still lack a unifying picture, despite sharing a common underlying mechanism.

Here our aim is to suggest  a general description of cooperative behavior for multi-sites systems by using a basic statistical mechanics approach, able to emphasize the role of interactions between sites and to recover all the different behaviors and definitions.
In our model, binding sites are associated to two-values variables, indicating the binding/non binding state analogously to the classical Ising model \cite{thompson}\cite{barra1}.

A first step in this direction has already been paved\cite{thompson} by using linear spin chains for modeling nearest neighbors correlations among sites; here we move forward, by assuming a simple mean field interaction structure among active sites as this allows to understand complex behaviors of macromolecules. More precisely, minimizing a suitably introduced free energy, which depends on proper order parameters, we obtain an explicit expression of the reaction rate as a function of substrate concentration and of a possible coupling among binding sites.
Within this framework, we can then clarify several general properties of cooperative binding, hence allowing a unified picture of cooperative behavior in reaction- rate curves. In particular, we show that the strength of the interaction among active sites discriminates in a simple way between Michaelis-Menten curves and cooperative behavior, the latter in agreement with several, well-known criteria (e.g. Hill, Koshland, range, etc.); furthermore we identify a critical interaction threshold where strong cooperativity sets in/out, also showing that a sigmoidal rate law is not necessary for cooperativity. In our framework, also reaction rates curves with an almost discontinuous step in a defined range of concentration
can be recovered. In particular, the behavior of the reaction curves around the discontinuous step can be given by an interesting representation in terms of "critical" parameters.

Finally, by extending our model to multi-sites attachment (a reasonable assumption when binding occurs on macromolecules or on a homogeneous surface), we obtain reactions curves exhibiting different features (weak, strong, sigmoidal or discontinuous step), as observed  experimentally \cite{cook}, and negative cooperativity may naturally arise.

The paper is organized as follows: In section $2$ we briefly summarize the principal definitions of cooperativity in chemical kinetics, and in section $3$ we introduce our statistical mechanics framework. The latter is then throughout investigated as explained in section $4$, where we also recover previous definitions of cooperativity. In section $5$ we properly extend our model in order to account for multi-attachment too. Outlooks and conclusions then follow.

\section{Measures of cooperativity}

As in elementary chemistry, cooperativity  can be defined as the ability of ligand binding at one site on a macromolecule to influence ligand binding at a different site on the same macromolecule; according to whether the affinity of the binding sites for a ligand is increased or decreased upon the occupation of a neighboring site, we speak of positive or negative cooperativity respectively. A common example is provided by hemoglobin, which displays four binding sites whose affinity for oxygen is increased when the first oxygen molecule binds \cite{thompson}.

In order to study the existence and the extent of cooperativity, a convenient
observable is the fractional saturation of binding sites at equilibrium $\theta $ ($\theta=$ occupied sites/ total sites) for different values of free ligand concentration $\alpha$ \footnote{In chemical kinetics one is typically interested in finding the rate law governing a reaction,
i.e. how the concentration of a given reactant or product varies in time.
For elementary reactions (i.e. reactions with a single mechanistic step)
the law of mass action is valid, and the reaction rate is proportional to reactant concentrations raised to a power defined by stoichiometric coefficients,
but this is no longer true for complex reactions.
However, if the substrate molecules react on binding sites to form a product, as is the case of enzymes,
we expect the rate for product formation to be still proportional to the fraction of occupied sites $\theta$.}.
Due to the number of possible forms and levels of cooperativity, a multitude of diagnostic tests have been introduced in literature (see for instance \cite{karlin}), mainly based on the comparison to a non-cooperative behavior.
The latter occurs when the binding sites (e.g. enzyme units) act independently and are identical in their activity; this behavior is observed for example in myoglobin in the presence of oxygen \cite{thompson}. In such cases, the equilibrium reaction curve is well defined and exhibits the familiar hyperbolic Michaelis-Menten (MM) form
\begin{equation}
\theta_M(\alpha) = \frac{\alpha}{\alpha+K_M},
\end{equation}
where $K_M$, called the MM constant, rules the affinity between components, and $\alpha$ accounts for the substrate concentration; note that $\theta_M$ correctly saturates to $1$ \cite{thill}.
Hence, cooperativity is expected to be at work when the reaction rate does not follow MM kinetics.

We now review different measures and, accordingly, definitions of cooperativity (see also \cite{karlin,bardsley,karlin2,acerenza}), neglecting those
(e.g. Wong co-operativity \cite{wong}), which are restricted to particular forms for the reaction rates.
We also omit those based on the shape of $\theta(\alpha)$ near the origin, as this is not necessarily indicative of the behavior at intermediate substrate concentrations: Indeed, mixtures of positive and negative co-operativity, different degrees of co-operativity and the occurrence of tipping points can not be predicted by the behavior of $\theta(\alpha)$ when $\alpha$ is close to zero.

\subsection{Koshland cooperativity}
The Koshland measure of cooperativity is one of the easiest to implement, being based on the coefficient $\kappa$, defined as the ratio
\begin{equation}
\kappa = \frac{\alpha_{0.9}}{\alpha_{0.1}},
\end{equation}
where $\alpha_{0.1}$ and $\alpha_{0.9}$ are defined as those values of ligand concentration such that $\theta(\alpha_{0.1})=0.1$ and, similarly, $\theta(\alpha_{0.9})=0.9$: For the MM hyperbolic function $\theta_{M}(\alpha)$, one would get $\kappa = 81$, independent of $K_M$. Hence, one can notice that if $\kappa <81$ then $\theta(\alpha)$ is expected to grow faster than $\theta_{M}(\alpha)$ and, according to Koshland, we speak of positive cooperativity, while if $\kappa > 81$, then we have negative cooperativity.
\newline
The index $\kappa$ has the advantage of being an absolute number, but the disadvantage of being sensitive to the scale of measurements and of depending only on $\alpha_{0.1}$ and on $\alpha_{0.9}$, hence ignoring all information that can be derived from the shape of $\theta(\alpha)$.

\subsection{Global dissociation quotient}
\label{sec:global}
Another possible approach to measure the degree of cooperativity is in terms of deviations from
the hyperbolic behavior by introducing the generalized equation \cite{edsall}
\begin{equation}\label{eq:quotient}
\theta(\alpha) = \frac{\alpha}{\alpha + K(\alpha)}.
\end{equation}
In the absence of cooperativity $K(\alpha)$ is constant, viceversa one gets $K(\alpha) = \alpha(1-\theta)/\theta$.
Hence, the experimental value of $K(\alpha)$ can be interpreted as the ratio between free sites and occupied sites times $\alpha$ and it is often called ``global dissociation quotient''.
The slope of this curve, symbolized by
$\gamma$, i.e.
$\gamma = dK(\alpha) / d \alpha$, is used as a quantitative measure of cooperativity:
$\gamma$ equal, greater or smaller than zero would correspond
to absence of cooperativity, negative and positive
cooperativity respectively. Under this view, the phenomenon
of cooperativity is equivalent to the change
in the global dissociation quotient with ligand concentration; the larger the rate of change of $K(\alpha)$ and the larger the degree of cooperativity.

\subsection{Hill cooperativity} \label{sec:hill}
An alternative way to quantify the deviation from
an hyperbolic behavior is to use the so called Hill function, which provides another extension of the MM function and which reads
\begin{equation} \label{eq:hill_function}
\theta_H(\alpha) = \frac{\alpha^h}{\alpha^h+K_M},
\end{equation}
with $K_M$ and $h$ constant. Indeed, Eq. (\ref{eq:hill_function}) is able to successfully fit most experimental data from cooperative systems, where the parameter $h$ can be interpreted as the number of ligand molecules that bind to the macromolecule in a single step \cite{hill}.
\newline
More generally, given experimental data for $\theta(\alpha)$, one can define the Hill coefficient as
\begin{equation} \label{eq:hill}
n_H(\alpha)=\frac{d \left[ \log \frac{\theta(\alpha)}{1 - \theta(\alpha)} \right]}{d (\log \alpha)} = \alpha \frac{d \left[ \log \frac{\theta(\alpha)}{1 - \theta(\alpha)} \right]}{d \alpha},
\end{equation}
which provides a measure of the extent of cooperativity: when $n_H (\alpha) >1$ for all $\alpha>0$, we have positive cooperation; viceversa, if  $n_H (\alpha)  \leq 1$ for all $\alpha>0$ we have negative cooperation.
For instance, for oxygen binding to hemoglobin (with $N=4$ binding sites), the estimated Hill coefficient is about $2.8$. Indeed, in biochemical processes $n_H$ is typically smaller than $N$, suggesting a mixed effect resulting from the interaction among sites \cite{cook}.

It is worth noticing that the main difference between $\gamma$ and $n_H$, is that the former
represents the absolute change in global dissociation
quotient per absolute change in ligand concentration,
while the latter is related to the corresponding
logarithmic changes. 

Finally, we underline that, given the experimental curve $\theta(\alpha)$, one can straightforwardly get an estimate for $n_H$, and this explains the popularity of such cooperativity measure, nonetheless $n_H$ constitutes a macroscopic property and it is not uniquely related to the difference in binding affinity due to the occupancy of next sites, which, conversely, is a microscopic property.

\subsection{Range cooperativity: weak or strong behavior}
The range measure is, again, based on a comparison between the reaction curve $\theta(\alpha)$ and an ``appropriate'' hyperbolic MM curve, as summarized in the following (see also \cite{karlin}):
first, one specifies a value $K >0$ and builds the hyperbolic function $\theta_{K}(\alpha) = \alpha /(\alpha + K)$. Then, if  $\theta'(0)< \theta'_{K}(0)$ ($\theta'(0)> \theta'_{K}(0)$) and $\theta(\alpha)$ and $\theta_{K}(\alpha)$ intersect at most once for all choices of $K>0$, we say that $\theta(\alpha)$ is \emph{strongly} positive (negative) cooperative.
If $\theta(\alpha)$ and $\theta_{K}(\alpha)$ intersect more than once, it is possible to formulate a concept of \emph{weak} cooperativity (in the range sense): If $\theta'(0)< \theta'_{K}(0)$ ($\theta'(0)> \theta'_{K}(0)$) and $\theta(\alpha) > \theta_{K}(\alpha)$ ($\theta(\alpha) < \theta_{K}(\alpha)$) for large enough $\alpha$, then $\theta(\alpha)$ is weakly positive (negative) cooperative.

As proved in \cite{karlin}, for monotone increasing rate curves the range and Hill measures of cooperativity are equivalent, more precisely, strong positive (negative) cooperativity with respect to the range measure is equivalent to a positive (negative) cooperativity according to Hill.
However, differently from Hill measure, which requires the estimate of the slope for $\log \theta(\alpha)$ against $\log \alpha$, range measurements do not require interpolating the reaction curve from the data, and accordingly, the diagnosis of cooperativity is relatively simpler and also offers additional advantages \cite{karlin}. In particular, the method is able to distinguish different degrees of cooperativity (weak and strong), even though such a distinction is still based on macroscopic evidences and cannot be directly ascribed to specific values or range of values in the microscopic parameters characterizing the phenomenon.

\section{Bridging chemical kinetics and statistical mechanics}
Let us consider a system (i.e. a big macromolecule, a homo-allosteric enzyme, or a catalyst surface) 
which has a set of $N$ interacting sites, numbered by an index $i=1, 2, \dots, N$.
Each site can bind one identical smaller molecule of a substrate, and we call $\alpha$
the concentration of free substrate molecules.
When a site has a molecule bound on it, binding on all the other sites
is enhanced (inhibited), which corresponds to a system with positive (negative) cooperativity.

In analogy with the Ising model (see for instance \cite{thompson}), we associate in complete generality to each binding site a dichotomic variable $\sigma_i$ which takes the value $+1$ if the $i$-th site is occupied,
and $-1$ if it is unoccupied.
A configuration of the molecule is then specified by the set of values $\sigma_1, \sigma_2, \dots, \sigma_N = \{ \sigma \}$.

Once this bridge has been established, we can speculate on the way the binding sites interact with the substrate and/or among themselves: as the interactions are based on elettromagnetic exchange forces the third law of dynamics holds and a  representation in terms of an Hamiltonian $H_N$ is allowed\footnote{Strictly speaking the intrinsic symmetry of the exchanges implies symmetry in the couplings which ensures detailed balance and the latter ensures the equilibrium canonical framework to hold, ultimately an Hamiltonian representation of the phenomenon.}.
In this way we are able to ask for free energy equilibria of this Hamiltonian, namely to ask for the minimum energy and the maximum entropy principles:
When the system is in equilibrium with a heat bath at a given temperature $T$, the probability of a configuration
with energy $H_N$ is proportional to the Boltzmann factor $\exp(-H_N/k_BT)$,
where $k_B$ is the Boltzmann constant and $T$ is the temperature or, more generally, the level of ``noise'' experienced by the system.

We first focus on the interaction between the substrate and the binding site which is expected to depend on both the substrate concentration and on the state of the binding site; we model such interaction by an "external field" $h$ meant as a proper measure for the concentration of free ligand molecules.
Then, for a non interacting model (whose potential cooperation features are obviously neglected) one can consider
a microscopic interaction energy given by
\be
\label{H_ni}
	H_N (\{ \sigma \}; h) = - h \sum_{i=1}^{N} \sigma_i.
\ee
We notice that for $h$ positive, $H_N$ is negative and large in absolute value whenever there is a prevalence of occupied sites, while it is positive if most sites are unoccupied.
On the contrary, when $h$ is negative the smaller, negative, values for this one-body interaction correspond to a prevalence
of unoccupied sites and vice versa.
So we can think at $h$ as the chemical potential for the binding of substrate molecules on sites\footnote{We stress that our formulation of the problem
allows to introduce concentrations as real variables (instead of a natural enumeration of the substrate elements), namely external fields, which has the mathematical advantage of retaining the canonical ensemble instead of the grand-canonical one.}:
When it is positive, molecules tend to bind to diminish energy, while when it is negative, bound molecules
tend to leave occupied sites.
If we assume that the free molecules are non interacting (which is strictly true when they are very diluted), the chemical potential can be expressed as the logarithm of the concentration of binding molecules
and one can assume that the concentration is proportional to the ratio of the probabilities of
having a site occupied with respect to that of having it empty \cite{thompson}.
In this simple case, being the sites
non-interacting, the probability of each configuration is the product of the single independent probabilities
of each site to be occupied and one finds
$$ \alpha \propto p(\sigma_i=+1)/p(\sigma_i=-1) = \exp(+h)/\exp(-h) $$
so that
\be
\label{halfa}
	h = \frac{1}{2} \log \alpha.
\ee
Hence the limit $h \to -\infty$ corresponds to a vanishing concentration $\alpha$, while when $h \to +\infty$ the concentration $\alpha$ goes to infinity as well.
It is straightforward to see that the reaction-rate associated to eq. (\ref{H_ni}) corresponds to the hyperbolic
Michaelis-Menten curve\cite{thompson}.

Once we have successfully tested  the statistical mechanics approach for the paradigmatic MM kinetics, we can proceed with our mapping and
account for cooperativity just by extending the one-body theory previously outlined:
To model the interaction between sites we simply introduce the full Curie-Weiss (CW) Hamiltonian \cite{ellis}
\be
\label{H}
H_N (\{ \sigma \}; J, h) = - \frac{J}{2 N} \sum_{i \neq j}^{1,N} \sigma_i \sigma_j - h \sum_{i=1}^{N} \sigma_i,
\ee
which, again, represents the microscopic energy of the system in a given configuration,
so that the lower $H_N$ and the larger the probability of finding that configuration. Of course, $J>0$ means that there exists a two-body interaction between the binding sites, and
the free energy is better minimized if sites ``cooperate'' by aligning consistently.

The Hamiltonian (\ref{H}) was originally introduced to describe a system of atoms with magnetic moments: the spins $\sigma$'s can point in two possible directions, $h$ represents an homogeneous external magnetic field acting on each atom and spanning all real numbers,
and $J$ is a fixed number representing the two-bodies interaction strength (the exchange term),
which for our model rules the cooperativity of the system.
In fact, the first term in (\ref{H}) is an interaction between the $N (N-1)/2$ couples of sites: if $J$ is positive, this energy is smaller  the larger the alignment displayed by the $\sigma$'s; In the following we focus on positive couplings $J_{ij}>0$.
The first term ranges from its minimum value $ - J (N-1)/2$, obtained in the extreme cases in which all sites are either all occupied or all empty\footnote{Notice that this symmetry is actually broken in the presence of an external field, namely of a sufficiently large substrate concentration $\alpha$ which favors positive (occupied) states.}, to its maximum value $0$, obtained at half saturation of the sites.
So this contribution to the energy is such that sites tend to behave coherently,
and the constant parameter $J$ fixes the scale of the microscopic binding energy.

It is important to notice that our model corresponds to a so-called mean-field approximation, where each site interacts
with all the remaining $N-1$ with the same strength $J$: this assumption implies the minimal tractable oversimplification to investigate complex behaviors.
Even more refined models describing for example the binding of molecules to semiflexible polymers, can be treated with a mean-field approximation in the
limit of long-range interactions\cite{diamant}.  

First of all, a stochastic approach like ours is meaningful only if the number of elements taken into account is large; in particular, our solution holds in the thermodynamic limit
(i.e. $N\to \infty$); this implies that the predictiveness of our model can be significant even for a finite size
folded macromolecule, built up by, say, $N \sim 1000$ binding sites: The simplest way to realize a scaling able to retain the physical/chemical significance of the phenomenon under investigation\footnote{In fact complex phenomena as a folding of a macromolecule or a sudden jump in the reaction rate curve could not be taken into account by modeling the system by a linear chain with only nearest neighbors interactions. The world of long range models -whose applicability we are testing here- seem to be more consistent, and eventually more realistic diluted systems may be studied in future works.} is by assuming the number of neighbors for any site to scale with $N$ and this is indeed consistent with the mean-field approach. In this way the statistical overlap among a global behavior
and a local one is strong in the large $N$ limit. Otherwise stated, the features of an infinite length molecule, whose
interaction are properly rescaled within a mean field approach, can be still comparable with a realistic one.
Furthermore, still among the benefits,  by varying only the ratio $J/T$, our model provides a robust picture of the different regimes known for cooperative systems; it is also mathematically tractable and it accounts for a discontinuous
rate law  which  linear models with nearest neighbors interactions cannot recover.

From the other side, close to criticality, its prediction (which are unsensitive to the underlying topology) may barely represent the real phenomenon, quantitatively. However, as we will see, they can suggest an interesting measure to be performed on nearly discontinuous regimes, in the "critical" region.

The model could be further extended by including, for instance, a given topology for binding sites \cite{fed,sw}, in such a way that the effective number of neighbors per site is smaller than $N$ (and may even remain finite in the thermodynamic limit),
or a coupling $J_{ij}$ depending on the couple of sites considered, say, scaling with their distance; the resulting mathematics would be more complicated, without any real qualitative change.
Thus, keeping the  framework as simple as possible for the sake of clearness, we introduce the standard statistical mechanics definitions for the CW model. The normalized probability for a configuration $\{ \sigma \}$ is then
\be
	P_N( \{\sigma\}; J, h, T) = Z_N^{-1} \exp \left(-H_N(\{ \sigma \}; J, h)/k_B T \right)
\ee
where the normalizing factor
\be
	Z_N(J, h, T) = \sum_{\{\sigma\}} \exp \left( - H(\{ \sigma \}; J, h)/k_B T \right)
\ee
is termed the partition function.
The number $n \{ \sigma \}$ of occupied sites can be computed as
\be
	n \{ \sigma \} = \sum_{i=1}^N \frac{1}{2} (1 + \sigma_i),
\ee
and the binding isotherm $\theta(\alpha)$ is reconstructed from the average fraction of occupied sites
$\langle n \rangle / N$ as a function of the concentration $\alpha$
\be
\theta = \frac{\langle n \rangle}{N}  =  \frac{1}{N} \sum_{\{ \sigma \}} n\{ \sigma \} P\{ \sigma \} =  \frac{1}{2} + \frac{1}{2} m(J, h)
\ee
where $ m(J, h) = \sum_i  \langle \sigma_i \rangle / N $ represents the average magnetization per site in mean-field CW model,
 the brackets $\langle . \rangle$ account for the Boltzmann average (i.e. $\langle . \rangle =$ $\sum_{\sigma} . \exp(-\beta H_N) / $ $\sum_{\sigma} \exp(-\beta H_N)$);  the dependence on $\alpha$ is obtained simply by substituting $h$ with $(1/2) \log \alpha $ (see Eq.~$7$).
Here $m$ represents the average fraction of occupied (or unoccupied) sites with respect to half saturation.
\newline
In the limit of large $N$, the average value $ \langle n \rangle / N  $ can be
obtained by minimizing, with respect to the parameter $\theta \in (0,1)$, the effective free energy \cite{ellis}
\be
\label{free}
F( \theta ,\alpha, T) = \sup_{|\theta|}\Big( - \frac{J}{2} (2 \theta - 1)^2 - \frac{1}{2} (1-\theta) \log(\alpha)  - T s(\theta)\Big). \ee
The (only direct) independence of this variational prescription by the coupling constant (as the sup is taken just on $\theta$) is the main strand we pave to obtain a unifying picture of the various catalogues of cooperativity in chemical kinetics: It holds whatever $J$.
\newline
The first two terms at the r.h.s. of eq. (\ref{free}) stand for the internal energy, which corresponds to
the Boltzmann average of the Hamiltonian $H(\{ \sigma \},J,h)$ expressed as a function of $\theta$ and $\alpha$,
while
\be
s(\theta) = -\theta \log(\theta) -(1-\theta) \log \left(1-\theta \right)
\ee
is the entropic term, whose weight is ruled by the temperature $T$.
For small temperatures, the most likely configurations are those
associated to small values of the internal energy.
As we explained above, according to ligand concentration, $H_N$ is smaller either when there is a prevalence
of occupied or empty sites, and it is maximum at half saturation.

We also notice that the entropic term
is connected to the logarithm of the
number of configurations associated to a given fraction $\theta$ of
occupied sites: It is maximized for fractions around $1/2$, which have a larger
number of configurations associated, and minimized for $ \theta = 0, \ \theta=1$,
corresponding to just one configuration and a vanishing entropy.
The minimum of $F$ is the optimal compromise between the minimization
of the effective internal energy and the maximization of entropy.
We expect that for large interactions $J$ (or chemical potentials) the energy term in (\ref{free}) is the leading contribution, and the optimal
fraction is ruled by this term, so that the sites tend to be in the same state and the binding isotherm displays a sigmoidal shape,
while for small values of the interaction strength
the leading term will be the entropic one, which at a fixed value of the chemical potential prefers disordered states,
i.e. states obtained by a large number of configurations, and pushes the rate law towards a MM form:
In this sense
the temperature can be thought of as a noise
because when it is high the system prefers to be in
a disordered state. However, large or small temperatures are defined with respect to
the interaction strength $J$  and in the following we are going to fix the temperature equal to $1$
and let the interaction strength vary to see all the possible regimes of binding isotherms arising from different values of $J$,
 as the latter rules cooperativity.

The minimum condition for (\ref{free}) with respect to the order parameter $\theta$
corresponds to the CW self-consistence equation \cite{barra1}\cite{ellis}
\be
\label{tetacw}
	\theta (J,\alpha) - \frac{1}{2} = \frac{1}{2} \tanh \left( J (2\theta-1) + \frac{1}{2} \log(\alpha) \right)
\ee
and gives the average fraction of occupied sites corresponding to
the equilibrium state for the system.
As we will see later, this equation describes a second order phase transition when $\alpha=1$ and $J$
larger than the critical value $J_c = 1$. Conversely, when $J > J_c$ and $\alpha = \alpha_c = 1$, the transition
is first order and the average fraction $\theta$ has a discontinuity.
Equation (\ref{tetacw}) holds rigorously  just in the thermodynamical limit ($N \to \infty $);
one could also consider corrections of order $(1/N)$ (see for instance \cite{ellis}), but for large systems these are not important, the only relevant
change is that at finite volume the discontinuous functions are mildly smoother (accordingly with the experimental counterparts \cite{thill}).

Summarizing, the binding isotherm is a continuous function of $\alpha$ for every fixed $J \leq 1$,
and it has a discontinuous jump in $\alpha=1$ for $J>1$. The value $J=1$ corresponds to a critical
behavior where the derivative of $\theta$ with respect to $\alpha$ becomes infinite at the reference
concentration $\alpha =1$, and in the following we will see that this is connected to an infinite Hill coefficient\footnote{An infinite Hill coefficient may seem unrealistic, however it is not an unavoidable feature of our modeling: in fact $h$ scales with the connectivity of the underlying network of interactions and, while the latter diverges in this minimal fully connected representation, diluted mean fields can still work fine. However, as the mathematics involved becomes immediately very heavy we preferred to present the pure theory within the limitation of the high connectivity limit.}.

\section{Obtaining chemical kinetics from statistical mechanics}

As discussed in the introduction, to determine the rate law one has to compute the average fraction of occupied sites as a function of the
concentration of free ligand molecules $\alpha$, and this dependence is encoded in the self-consistence equation (\ref{tetacw}).
\newline
We consider separately the two cases $0 \leq J <1$  and $J>1$ for the interactions between sites,
because, as stated, while in the former case $\theta$ is everywhere continuous in $\alpha$, in the latter it has a discontinuity
in $\alpha =1$: it takes a value smaller than $1/2$
when $\alpha \to 1^-$ and greater than $1/2$ when $\alpha \to 1^+$\footnote{In the magnetic counterpart this corresponds
to the fact that below the critical temperature we have a ferromagnetic phase with a positive remanent magnetization when $h \to 0^+$,
and a negative one when $h \to 0^-$.}.
In either cases, however, it is easy to check that $\theta(\alpha) \to 0$ for $\alpha \to 0$ (which corresponds to the $h \to -\infty$ limit)
and $\theta(\alpha) \to 1$ for $\alpha \to \infty$ ($h \to +\infty$). This means that the
reaction rate vanishes when the substrate concentration vanishes and it saturates to one
(all the binding sites being occupied) when the substrate concentration is large, as expected.

\begin{figure}[htb!]
  \includegraphics[width=10cm]{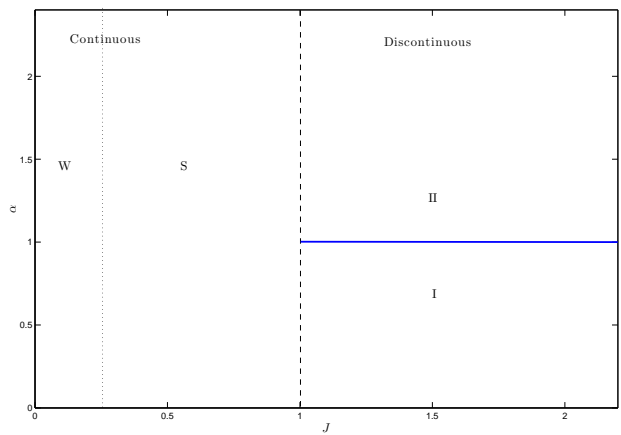} \\
  \includegraphics[width=10cm]{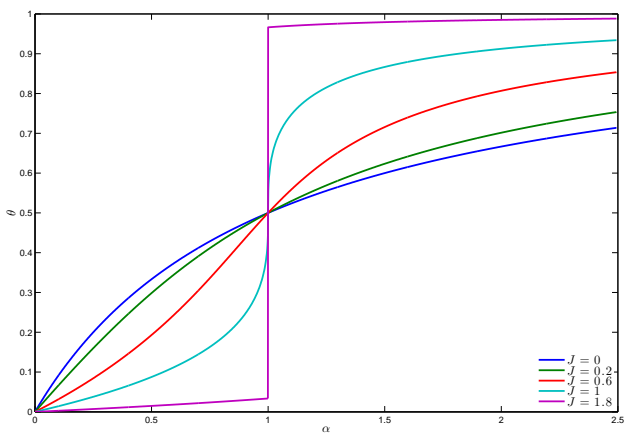}
   \caption{Top: \
	the figure shows the different phases for the system. 
	For $J<1/4$ the binding isotherms have no inflection point and the system is weakly (W) cooperative;
	when $1/4<J<1$ an inflection point arises, 
	and the presence of a sigmoidal shape allows us to call this regime strongly (S) cooperative;
	both regions correspond to a continuous varying $\theta(\alpha)$, 
	while for $J>1$ $\theta(\alpha)$ becomes discontinuous,
	with a jump at the critical concentration $\alpha=1$(blue line).
	Bottom: the different regimes for the binding isotherm obtained by varying
	the interaction strength  $J$. For $J=0$ (blue line) the hyperbolic Michaelis-Menten law represents the isotherm
	for a non-interacting system;  for $J = 0.2$ (green) the system is in a weakly cooperative regime;
	for $J=0.6$ (red) strong cooperativity manifests itself with the typical sigmoidal shape;
	$J=1$(light green) is the critical regime: the derivative in the inflection point which gives the Hill coefficient is infinite;
	$J=1.8$ (purple) represents the discontinuous phase, with an extremely strong cooperativity.
  	\label{fig:fasip2}}
\end{figure}

When $J \to 0$, no cooperativity is expected (as the model reduces to a one-body theory) and, coherently, we recover the MM kinetics.
\newline
In fact the rate (\ref{tetacw}) can be equivalently expressed as
\be
\label{f2}
	\theta(\alpha, J) = \frac{\alpha \exp \left[ 2J (2 \theta (\alpha, J) - 1)  \right] }{1 + \alpha \exp \left[  2J  (2 \theta( \alpha, J)-1)  \right]}
\ee
which properly gives, for $J=0$
\be
\label{f0}
	\theta(\alpha, J)|_{J=0} = \frac{\alpha}{1+ \alpha}.
\ee
Note that we do not lose generality when obtaining $\alpha/(1+ \alpha)$ instead of $\alpha/(K+ \alpha)$  at the r.h.s. of eq. (\ref{f0}):
in fact we can rewrite the MM equation as $\theta(\alpha)=K_M^{-1}\alpha/(1+K_M^{-1}\alpha)$ such that choosing $K_M \neq 1$ is equivalent to shifting
$\alpha \to \alpha/K_M$, which can be compensated by shifting even $h \to (1/2)\log \alpha - (1/2)\log K_M$.
\newline
Moreover, from eq. (\ref{f2}), we get that when $J>0$ the rate for a given concentration is smaller than the
corresponding one for a non interacting system when $\alpha <1$, and becomes greater when $\alpha>1$, in fact,
the greater the interaction and the steeper the sigmoidal shape of the  the rate.
Equation \ref{f2} is plotted in Figure \ref{fig:fasip2} versus $\alpha$ and for several values of $J$.

\subsection{Continuous rate-law}
As we said before, for $J<1$, $\theta(\alpha, J)$ is everywhere continuos\footnote{For the magnetic model this is a paramagnetic phase,
meaning that there is no remanent magnetization when the external field vanishes: $m(\alpha, J)|_{\alpha=1} = 0$.}
and, when the concentration is equal to the reference concentration $\alpha = 1$, one half of the sites is  occupied in average.
The shape of the kinetics law, however, depends on the interaction between sites.
The derivative\footnote{Note that in the frame of the Curie-Weiss model this is
strictly related to the generalized susceptibility $$ \chi = \frac{\partial m(h)}{\partial h} $$ which measures the response
of the system to a change in the field $h$. In fact, we have
$$ \frac{\partial \theta}{\partial \alpha} = \frac{1}{2}\frac{\partial m(h(\alpha))}{ \partial \alpha}
= \frac{1}{2}\frac{\partial h}{\partial \alpha} \chi(h(\alpha)) = \frac{1}{4 \alpha} \chi(h(\alpha))$$
}
of $\theta$ with respect to $\alpha$, which is strictly related to the Hill coefficient and, consequently, to the cooperativity of the system,
can be computed from (\ref{tetacw}):
\be
\frac{\partial \theta}{\partial \alpha}    =  \frac{1}{4 \alpha} \frac{1- (2 \theta -1)^2}{1 - J \left[ 1-(2 \theta -1)^2\right]}.
\ee
It is always positive and finite for $J<1$, meaning that $\theta$ is an increasing function of $\alpha$,
as we expect. In the limit of low concentration we obtain
\be
\label{df}
\left. \frac{\partial \theta}{\partial \alpha} \right|_{\alpha=0}= \exp (-2 J)
\ee
and the kinetics at very low concentration is governed by the two-bodies interaction $J$:
the greater $J$ and the flatter the rate law. Note that the strength of the cooperativity, $J$,
appears in eq. (\ref{df}) as an exponent, implicitly supporting the log-scale of the Hill coefficient in eq. $(5)$.
When $J=0$, $\partial_{\alpha} \theta |_{\alpha=0}=1$ and one properly recovers the same trend as that of the MM kinetics,
which has a first order kinetics with the same coefficient for small concentrations.

\subsection{Weak and strong cooperativity}
The second derivative can be computed and expressed in terms of the first one:
\be
\label{ddf}
	\frac{\partial^2 \theta}{\partial \alpha^2} = - \frac{1}{\alpha} \frac{\partial \theta}{\partial \alpha}
	\left[ 1 + \frac{2 \theta -1}{\left( 1 - J\left[ 1-(2 \theta -1)^2\right] \right)^2} \right].
\ee
When $\alpha$ ranges in $(1, \infty)$, $m$ is positive and the second derivative is negative, so that $\theta$ is a concave function of $\alpha$
in that range. For $\alpha=1$ we have $\partial_{\alpha^2}^2 f = -(1/4)/(1-J)$, so that $\theta$ is still concave there.
For $\alpha \in (0,1)$ a numerical study of the second derivative clearly shows that it vanishes for a unique $\alpha^*$ depending on $J$.
This $\alpha^*$ is near to $\alpha=1$ for values of the interaction close to $J=1$, and it reduces with $J$ towards $\alpha = 0$.
Interestingly, the inflection point does not disappear when $J=0$, but for a greater value, namely $J=1/4$.
In fact, expanding  $m$ in the first order in $\alpha$ one finds
\be
\label{ddf0}
	\left. \frac{\partial^2 \theta}{\partial \alpha^2} \right|_{\alpha=0} = - 2 (1-4J)\exp(-4J)
\ee
so, when $1/4 < J <1$, there is a unique inflection point $\alpha^*$ depending on $J$,  which
separates the region where $\theta$ is convex (small concentration), to the one where it is concave.
For $J=1/4$ the inflection point corresponds to $\alpha^* = 0$ and when $J < 1/4$ the rate function $\theta$
is everywhere concave, tending, for $J \to 0$, to the hyperbolic MM form (whose second derivative $-2(1 + \alpha)^{-3}$
is always negative). Note that when $J=0$ the expression (\ref{ddf0}) gives, correctly, the MM value $-2$.
The absence of an inflection point in the region $J \in [0, 1/4] $ allows us to define it as a weak cooperativity region,
while we can call strong cooperativity the one arising in the interval $ J \in [1/4, 1]$.

These very simple definitions
have the advantage of being directly related to the microscopic interaction, so that the experimental behavior
of a system could allow one to reconstruct this interaction strength and interpret the rate law in terms of
the very general mean-field model we introduced.
\newline
Summarizing, we saw that the rate is continuous in $\alpha$ in the whole range $0<J<1$, and it has
a unique inflection point in the region $1/4<J<1$, and no inflection points for $J<1/4$ (see Figure \ref{fig:fasip2}).
As a sigmoidal rate has necessarily an inflection point, we talk about strong cooperativity in the former case and weak cooperativity in the latter.

\subsection{Strong cooperativity and discontinuous rate-law}
\label{sec:strong}

The different phases for the binding isotherm are shown in figure Figure \ref{fig:fasip2}.
When $J<1$ we can consider two regions, a weak cooperative one, where the rate law is hyperbolic,
and a cooperative one with an inflection point for the rate law that grows gradually with the interaction strength.
\newline
When $J >1$ (corresponding in the original Ising model to the "ferromagnetic" phase) the rate law is still increasing with $\alpha$, and the expressions (\ref{df}-\ref{ddf0}) remain valid for $\alpha \neq 1$.
In this point the rate function is discontinuous and the jump is given by $ \theta_+(J)-\theta_-(J) $,
where $$ \theta_{\pm}(J) = \lim_{\alpha \to 1^{\pm}} \theta(\alpha, J). $$
These two limits depend on $J$: they are both equal to $1/2$ for $J=1$, when the curve
is still continuous, and their difference increases smoothly with the square root of $J-1$ when
$J>1$.
This means that, starting from vanishing concentration, the system has less sites occupied, for a given $\alpha$, than the corresponding non interacting one,
until the concentration reaches the reference value. Here, it is sufficient to increase infinitesimally the number of free molecules to obtain a large filling
(depending on $J$). After that value, the number of occupied sites is always greater than the corresponding value for MM.
Note that, in principle, if the concentration varies slowly one could observe metastability, with a curve
which continues growing continuously up to values of $\alpha >1$. The entire out of equilibrium features of the model are ruled out in this treatment as we deal with equilibrium statistical mechanics, however -as a second step- the bridge could be extended in that direction.
If $J \to \infty$ this discontinuity increases, while its derivative in zero vanishes, so that in the large volume limit we obtain a step function.
This corresponds to a chemical kinetics where no binding site is occupied until the concentration has reached the critical value $\alpha =1$,
and when this value has been reached all sites are occupied.
This kind of discontinuous behavior can be observed, for example, in the binding isotherms of small surfactants onto a polymer gel \cite{murase}.

When $J \to 1$ a second order phase transition appears. This indicates that the correlation between binding sites becomes stronger  and the typical trend of thermodynamical observables is a power law \cite{ellis}. If the statistical mechanics picture we suggest applies to the reaction rates, then this power law behavior becomes very interesting, as it should be related, as in the statistical mechanics counterpart, to very general features the systems, due to the concept of universality \cite{ellis}. Let us consider in details the behavior  for $J \to 1$.
As we said before, the discontinuity for $J>1$ is given by $ \theta_+(J)-\theta_-(J) $, whose dependence from $J$ near the critical point $(\alpha=1, J=1)$
can be expressed as
$$  \theta_+(J)-\theta_-(J) \approx (J-1)^{1/2}$$
while on the critical isotherm (i.e. for $J=1$) around the critical concentration $\alpha=1$, mean-field theory predicts
$$ \theta (\alpha, 1) -\frac{1}{2} \approx \left( \alpha - 1\right)^{1/3}. $$
In this regime one can also predict the behavior of the $\alpha$ dependence of the Hill coefficient defined in (\ref{eq:hill}),
which, when $\alpha \to 1$, diverges as
$$ n_H (\alpha) \approx \left( \alpha - 1\right)^{-2/3}. $$
Moreover, we know that when $J \to 1^{\pm}$ the susceptibility, and so the derivative of $\theta$ respect to $\alpha$, diverges as
$$ \chi|_{\alpha =1} \approx \left| 1- J \right|^{-1}. $$
As hinted in Sec. $3$, the exponents of these power laws are only valid in the limiting case of a very large number of interacting neighbors per site,
and we expect that the right exponents should depend on the real dimension of the space in which they are embedded, as in the corresponding statistical mechanics models. However these scalings, in particular those related to the reaction rate around the discontinuity as a function of $\alpha$, suggest a new interpretation of almost discontinuous reaction curves and could represent an interesting measurements to test our predictions. Indeed, statistical mechanics suggests that power-law behaviors in the vicinity of the discontinuous step are signatures of a critical phenomena:
If in particular the detected exponents do not differ that much from the ones predicted by the CW model,
then this suggests that the mean field picture, despite its simplicity, is able to capture the essence of the cooperative phenomenon.

\subsection{Cooperativity through the Hill coefficient}

As explained in Sec.~\ref{sec:hill}, the usual way to define in a quantitative manner the cooperativity of a system is by the Hill coefficient,
i.e. the slope at the symmetric \footnote{
When there is not such a symmetry, one should more properly consider the quantity
\be
n_H = 4 \left. \frac{\partial \theta}{\partial \log \alpha}\right|_{\mbox{max, min}}
\ee
that is the slope of $\theta(\log \alpha)$ calculated at the inflection point, where it has an extremum.
In fact for an unsymmetrical system the slope at $\theta = 1/2 $ is not generally an extremum slope and
should not be taken as a suitable cooperativity index.}
point $\theta = 1/2$
\be
n_H = \left. \frac{1}{ \theta (1-\theta)} \frac{\partial \theta}{\partial \log \alpha} \right|_{\theta = 1/2} = 4 \left. \frac{\partial \theta}{\partial \log \alpha}\right|_{\theta = 1/2}
\ee
If binding on different sites is an independent process, one simply finds $ n_H = 1 $, while in the extremum case
in which sites are either all empty or all occupied $ n_H = N $.
We call a system cooperative (non cooperative) if $n_H>1 \; (n_H =1)$, while the cooperativity is said to be negative, meaning
that binding is reduced if there are occupied sites, for $n_H<1$. So this gives a lower bound for the number of interacting sites,
and it is possible to see that it is related to the variance of the mean number of occupied sites.

The Hill coefficient for our general model depends, as expected, on the interaction $J$;
in particular for $J<1$ we have
\be
n_H \equiv \frac{\partial \log \left[ \theta / (1-\theta) \right] } {\partial \log \alpha}
  = 4  \left. \frac{\partial \theta}{\partial \alpha} \right|_{\alpha = 1} = 1/(1-J)
\ee
Being the derivative of $\theta$ for $\alpha=1$, the Hill coefficient is finite (and greater than one) for $J<1$ and it diverges for $J \to 1^-$
when the discontinuity appears.
We observe that the last equation relates $n_H$ to the typical scale of interaction energy.
This is a macroscopic measure of cooperativity which is directly associated to the microscopic interactions among sites.

\subsection{Cooperativity through the global dissociation quotient measure}

\begin{figure}[htb]
	\includegraphics[width=10cm]{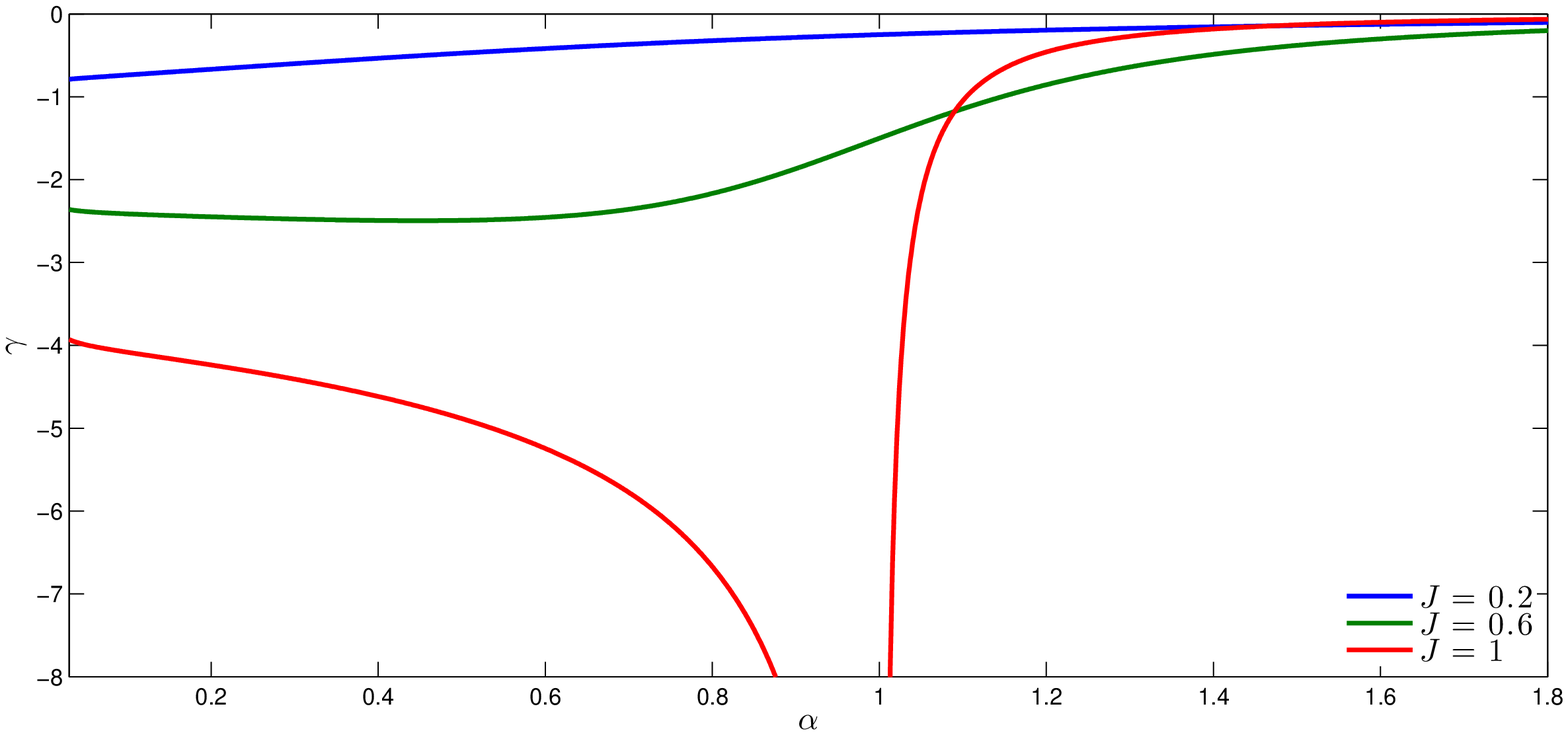} \\ \includegraphics[width=10cm]{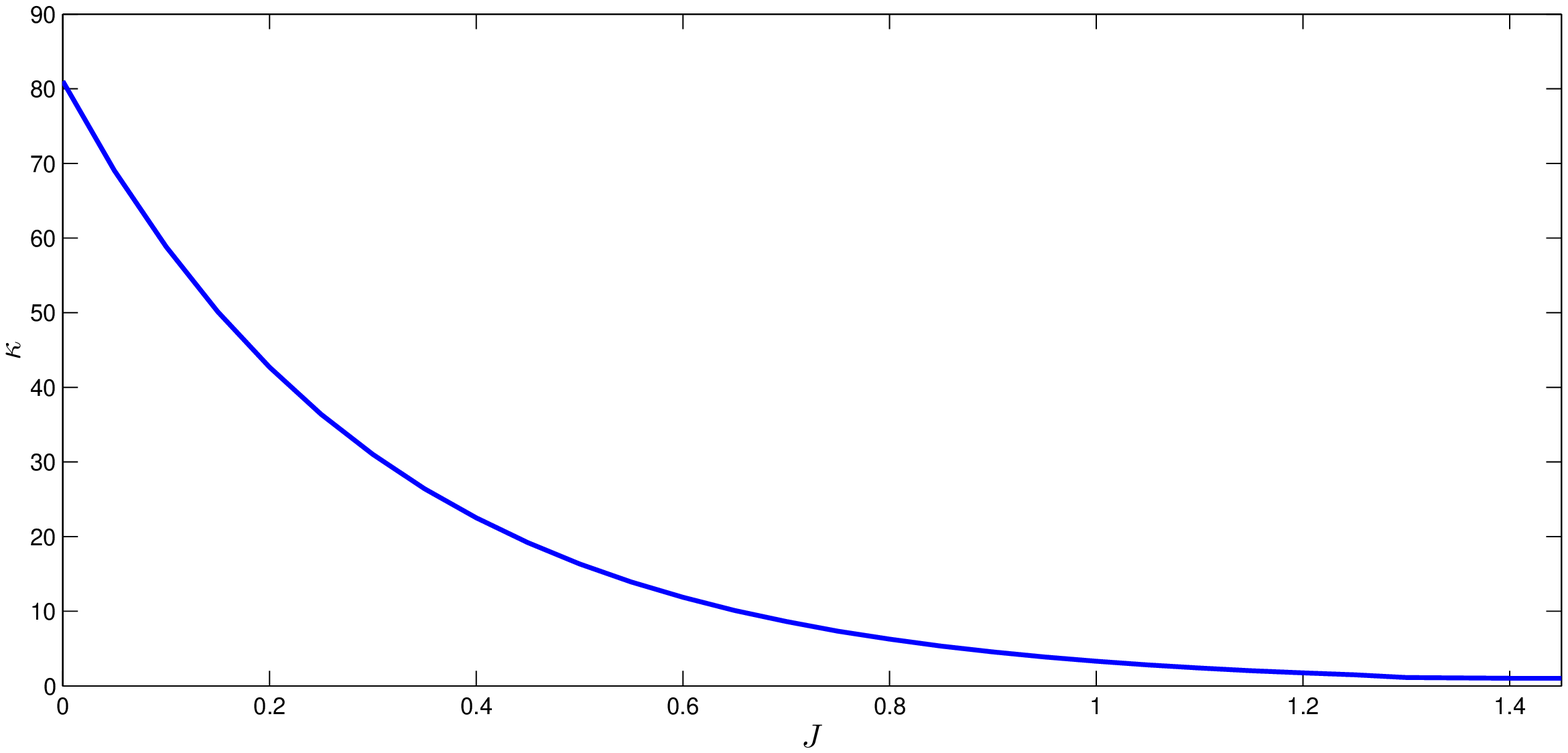}
	\caption{Top. \ The figure shows different regimes for the derivative of the global dissociation quotient $\gamma(\alpha)= dK/d\alpha$ obtained by varying
	the interaction strength  $J$. For $J = 0.2$ (green) $\gamma$ remains quite close to zero, indicating that $K$ is nearly constant in $\alpha$, as
	one expects for a non-cooperative system; for $J=0.6$ (red) the global dissociation quotient has a stronger dependence on $\alpha$: we know that in this case
	the binding isotherm has a sigmoidal shape and strong cooperativity appears; when
	$J=1$(cyan) we are in the critical regime, where the derivative of $\theta$, and so $\gamma$, diverges at $\alpha=1$.
  Bottom. \  The Koshland coefficient $\kappa=\alpha_{0.9}/\alpha_{0.1}$ obtained by our theory as a function of the two-bodies interaction $J$.
    It is a decreasing hyperbolic function, defined for values of $J$ below $J=1.5$, which tends to be lower when cooperatively is larger
    and the binding isotherm takes a steeper sigmoidal form.
	  	\label{fig:global}}
\end{figure}

Among the useful tools to describe cooperativity, we recall the global dissociation quotient $K(\alpha)$, whose derivative $\gamma$
is expected to be different from zero when some form of cooperation occurs; in particular, $\gamma$ should be negative when there is a positive cooperation. Figure \ref{fig:global} shows some plots of $\gamma(\alpha)$ for different values of $J$: consistently with the definition of $K(\alpha)$, (see \ref{sec:global}), in the region of $\alpha <1$, the stronger the interaction, the smaller  $\gamma$, properly indicating the deviation
from the $K$-constant MM curve. As the derivative of $\theta$ with  respect to $\alpha$ appears in the definition of $\gamma$, in correspondence of the critical value $J=1$, $\gamma(\alpha)$ diverges for $\alpha=1$. In fact this is the (critical) region where the correlations between sites, and so the cooperativity, are expected to diverge.

For sake of clarity in the figure, we did not show a plot of $\gamma$ for $J>1$, however
in this case the curve is below the others shown for $\alpha<1$ and it is not defined for $\alpha=1$: from this point on, it is practically vanishing.

\subsection{Cooperativity through the Koshland measure}

Lastly, we want to recover even the measure of cooperativity introduced by Koshland (see sec. $2.1$):
To satisfy this task, we plotted the previously defined Koshland coefficient  as a function of the two-bodies interaction coupling
$J$ (see Figure \ref{fig:global}), as the latter is the only relevant tunable parameter to explore cooperativity at work.
\newline
The coefficient is defined only for $J<1.5$; in fact for $J>1$ $\theta(\alpha)$ is a discontinuous function and when $J \geq 1.5 $ the equations
$\theta(\alpha)=0.1$ and $\theta(\alpha)=0.9$ cease to have a solution.
\newline
As shown in Figure \ref{fig:global} (bottom), we find that it is a decreasing function of $J$, which takes the value $\kappa=81$ for non-interacting systems ($J=0$), and $\kappa \sim 1$ when $J=1.4$:
Coherently with his scenario, we note that $\theta(\alpha = 1) = 1/2$ for every $J\leq 1$, when $J>0$ the binding isotherm is below the $MM$ isotherm for $\alpha < 1$ (such that $\alpha_{0.1}$ is larger than in MM theory) and above MM curve for $\alpha > 1$ (such that $\alpha_{0.9}$ is smaller than in MM theory).

\section{Multiple interacting systems}

When considering the binding of systems like long chain molecules, possibly on a homogeneous surface \cite{chat},
it may be possible that the interaction involves more than two hosting sites per time and this can in principle affect the global ''cooperative capability" of the system. Thus, in the following, we want to extend our model by accounting for multiple interactions, involving more than two elements together, in particular three and four bodies interactions.

The simplest $p$-body interaction among N sites which can be either occupied or empty ($\sigma_i = +1$ or $-1$ respectively) can be expressed by the Hamiltonian (\ref{pspin})
\be\label{pspin}
H_{N}(\{ \sigma \}, J, h) = -\sum_{p=2}^{\infty} \frac{p!}{2 N^{p-1}} J_p \sum_{1\leq i_1 < ...< i_p \leq N} \sigma_{i_1} ... \sigma_{i_N} - h \sum_{i=1}^N\sigma_i.
\ee
The interaction strength $J$ is assumed to be positive (for $J=0$ we recover the standard MM kinetics again) and the combinatorial factor before the summation makes the energy extensive and accounts for a sensible $p\to \infty$ limit \cite{barraPspin}.

It is easy to see that the energy (the mean value of the Hamiltonian) scales as
\begin{equation}\label{taylor}
\langle H_{N} \rangle \propto N \langle m  \Phi(m)\rangle, \ \Phi(m) = \sum_{k=1}^{p-1} c_k m^k +h,
\end{equation}
where with $c_k$ we meant the coefficient of the Taylor series implicitly defined by eq. (\ref{taylor}), and $m=2\theta-1$.

The argument in the self-consistent equation $m = \tanh(\Phi)$ is then built up (in complete generality) by all the terms of the Taylor series (whose convergence is tackle by the $p!$ term in the numerator).

However, as $|m|\leq 1$ we expect that only the first terms do matters in the thermal average and the main contribution is given by the $p=2$, which has been studied in details so far.

However, let us highlight here another advantage in performing the statistical mechanics approach: As  the Hamiltonian $\ref{pspin}$ represents a sum of Hamiltonians (each for each $p$), which, in principle, can have coupling strengths very different (i.e. the distribution of $J_p$ can range over several orders of magnitude), in this case solving the global problem as a whole can be prohibitive, while here we can consider each Hamiltonian contribution by itself and focus only on one term every time (namely $p=2,3,4,....$):

This procedure, which we are going to perform, allow to isolate the different contribution (and weight in the global behavior) of each independent binding capability represented by each independent $p$-term in the sum.

Note that the $\sigma's$ acting on a single site could tend to keep it in the state $-1$ or $+1$ in accordance to their product.
So, for example, for $p=4$ one could have three sites, acting on
the $i$-th site, in a configuration, say (+1, +1, -1), which favors the state $\sigma_i = -1$, even if in the whole there are more sites in the state $\sigma = +1$. The same local field is obtained when the configuration is (-1,-1,-1), i.e.
if all the three sites are empty.

The energetic behavior of this extension is then deeply different from the previous case: Usually one deals with linear forces, which are generated by quadratic potentials (i.e. $p=2$) such that the Maxwell-Boltzmann probability distribution is Gaussian accordingly with what naively expected by a simple Central Limit Theorem argument and critical behavior (at $h=0$) arises to confirm this picture \cite{barra1}.

However if a more complex scenario appears then a violation to this linear framework is expected: for instance for $p=3$ it is straightforward to check that even a positive $J$ may have subtle anti-cooperation features: in fact it is immediate to check that the energy would prefer the orientation of the spins $+1,+1,+1$ but also $+1,-1,-1$  (this can be thought of as a competitive feature of the multi-attachment that naturally introduce negative cooperativity in the process under investigation).

Interestingly, this extension still shares with the simplest $p=2$ case the same entropy: In fact,
as in the two-body case, for this long range interacting system the energy can be easily expressed
as a function of the parameter $\theta$ describing the fraction of occupied sites, and of the concentration $\alpha$
\be
\label{energyp}
N^{-1} H_p (\theta, \alpha) = -\frac{J}{2} (2 \theta - 1)^p - \frac{1}{2} \log(\alpha) (2 \theta - 1).
\ee
while the entropy per site is exactly the same as in the $p=2$ case. 
Given the effective free energy  (still keeping $T=1$ for the sake of simplicity)
\be
\label{freep}
F(\theta,J,\alpha) = \sup_{|\theta|}\Big( - \frac{J}{2} (2 \theta - 1)^p - \frac{1}{2}\log(\alpha) (2 \theta - 1)  - s(\theta)\Big),
\ee
the minimum condition with respect to the order parameter $\theta$  reads off as
\be
	\theta (J,\alpha) - \frac{1}{2}= \frac{1}{2} \tanh \left( \frac{1}{2} p J (2 \theta - 1)^{p-1} + \frac{1}{2} \log(\alpha) \right).
\ee
This corresponds to the equilibrium state for the system, and the average fraction of occupied sites
will be given by the solution of this equation. Once again,  this equation, as in the $p=2 $ case, is strictly valid when the number of sites is large.

Again, we expect that for large interactions $J$ (or chemical potentials) the energy term in (\ref{freep}) is the leading one, and
the sites tends to be in the same state (this corresponds to a large magnetization), while for small values of the interaction strength
the leading term is the entropic one, which prefers disordered states, i.e. states where the sites do not see each other.

\subsection{Case $p=3$}

\begin{figure}[htb]	
\includegraphics[width=10cm]{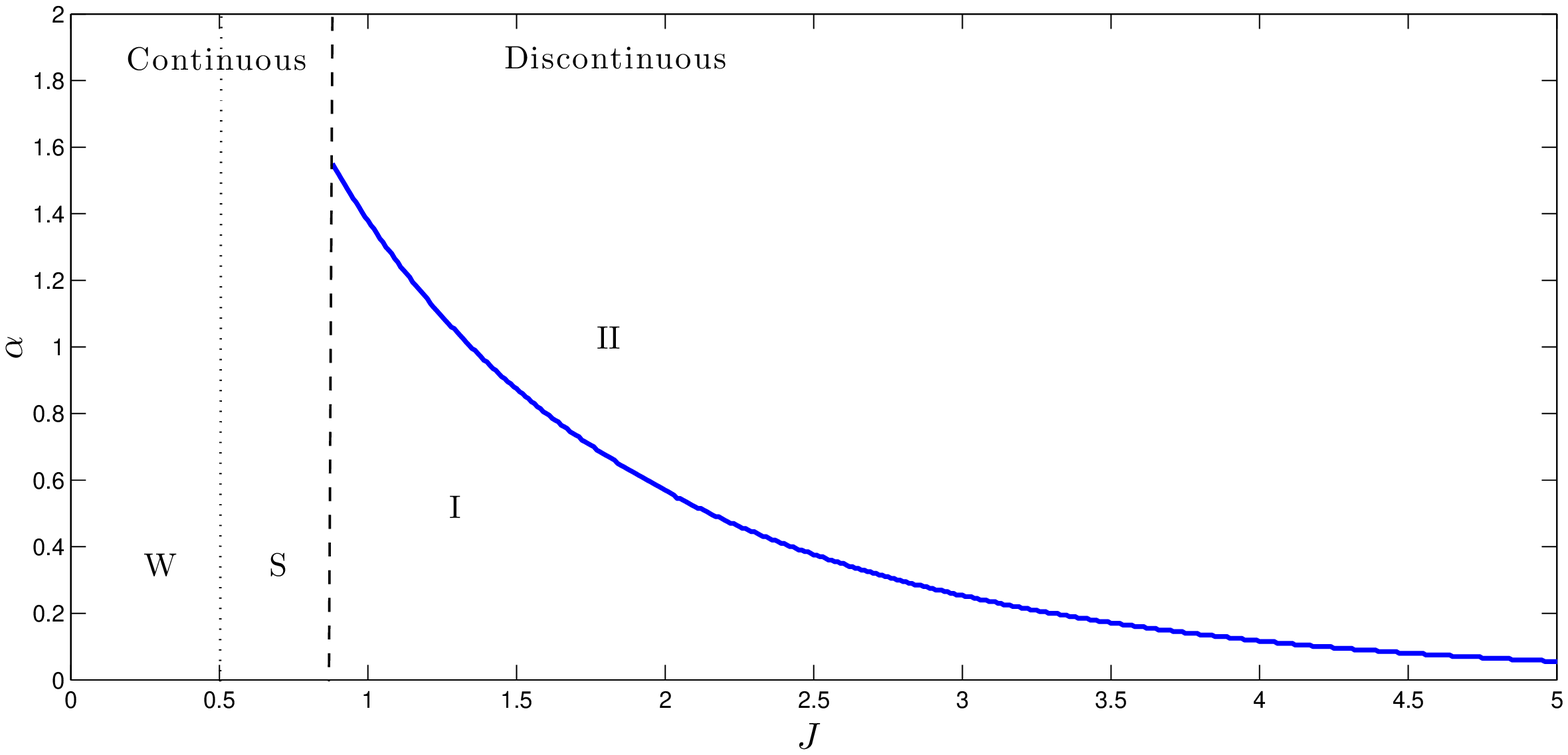} \\ \includegraphics[width=10cm]{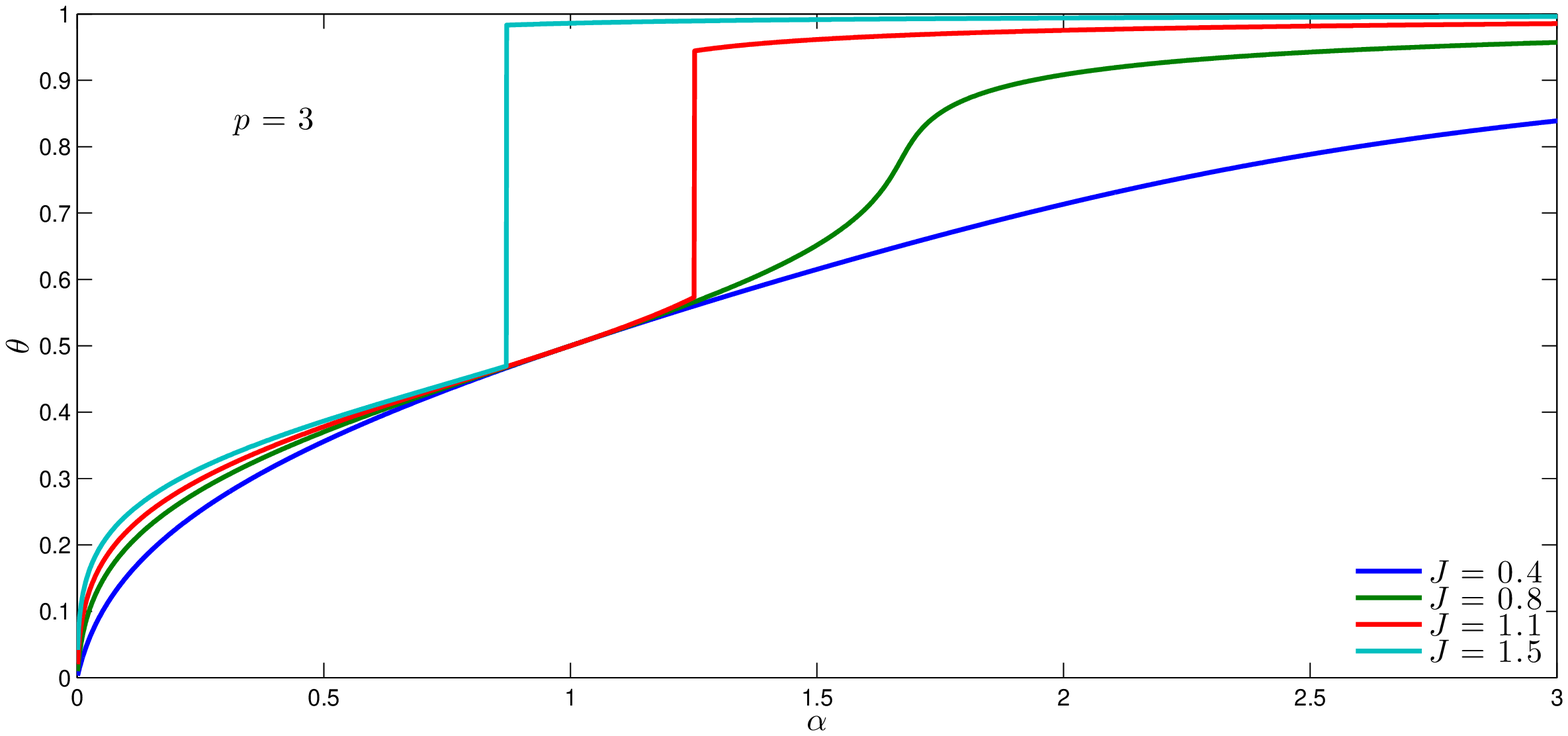}
\caption{$p=3$. Top: the figure shows the phases of a $p=3$ system. 
For $J_3<0.50$ the system is weakly (W) cooperative, with no inflections;
for $0.50<J_3<0.86=J_{3c}$ it is strongly (S) cooperative, having at least one inflection point;
the blue line shows the $J_3$-dependent critical concentration at which, for $J_3>J_{3c} $, a discontinuity in $\theta(\alpha)$
appears, passing from a relatively small $\theta$(I) to a larger one (II) as the concentration increases (at fixed $J_3$).
Bottom: binding isotherms $\theta(\alpha)$ for different interaction strengths $J_3$.
For $J_3=0.4$ (blue line) the system is weakly cooperative, with a hyperbolic form;
for $J_3=0.8$ (green) it has two inflections, the first for $\alpha=1/2$ and the second
for a larger $\alpha$; for $J_3=1.1$ (red) the isotherm has a strong inflection for $\alpha <1$
and a discontinuity for a value of the concentration between $1$ and $1.5$, corresponding
to a transition to a more ordered state; for larger $J_3$, as
the cyan curve for $J_3=1.5$ shows, the discontinuity is shifted towards $\alpha <1$.
Note that all the curves pass through the point $(1, 1/2)$.
 \label{fig:p3}}
\end{figure}

In this case the energy (\ref{energyp}) is an odd function of $m = 2\theta - 1$ at fixed $\alpha$
\be
N^{-1} H(\theta ,\alpha) = - J_3 (2 \theta - 1)^3/2 - \frac{1}{2}\log(\alpha) (2 \theta - 1).
\ee
Its global minimum, corresponding to a fraction of occupied sites $\theta$ and satisfying
\be
	 \theta = \frac{1}{2} + \frac{1}{2} \tanh \left( \frac{3}{2} J_3 (2 \theta -1)^{2} + \frac{1}{2} \log(\alpha) \right),
\ee
is located at  an average number of occupied sites smaller than $1/2$, when the concentration is very small.
On the contrary, when the concentration is around $\alpha=1$ (so that the chemical potential for binding is small)
and $ J_3 \gg 1$ the global minimum of energy corresponds to a large number of occupied sites.
There is an intermediate region where $H(\theta, \alpha)$ can have two local minima for $\theta$ smaller
and larger than $1/2$ respectively, but the relative strength of the chemical potential $\log(\alpha)$ versus the interaction $J_3$
tells us which is the global one, corresponding to the equilibrium state.
It is easy to see that the two minima (located respectively around $\theta = 1/4$ and $\theta=1$)
correspond to the same energy on the line $ \alpha = \exp (- 3 J_3 /4) $
(this is the critical line where there is a two phases coexistence in the limit of large $J_3$  or $\log(\alpha)$ and of negligible entropy contribution ($T\to0$)).
However the meaningful function to minimize is the whole free energy and in general one has to take into account the entropy term, which tends to
favor an equal number of occupied or empty sites and to disfavor deviation from this.

So we can identify three regions in the $(J_3,\alpha)$ plane: When $J_3$ is very small, the leading interaction is guided by the chemical potential  $\log(\alpha)$,
so that the sign of  $ \theta - 1/2 $ is the same as the former and it vanishes for $\alpha = 1 $.
This corresponds to a weakly(W) cooperative system.
Growing $J_3$, the system begins to feel mutual interactions, and the average number of occupied sites
as a function of $\alpha$ has an inflection point for $\alpha^*>1$, increasing quickly before that point
and more slowly after (see for instance the case $J_3 = 0.8$ in Figure $\ref{fig:p4}$). This is due to the interplay between entropy, which pulls the system
towards $\theta = 1/2$, and the energy, which prefers $\theta = 0$ or $\theta =1$ when the chemical potential is sufficiently large  (Figure \ref{fig:p4}).

As in the previous $p=2$ case, we can perform a numerical study of the second derivative with respect to $\alpha$ of the binding isotherm,
in order to discriminate between weak and strong cooperativity,  to understand when an inflection point, and consequently a sigmoidal shape, appears:
While for $p=2$ we found that this happens for $J=1/4$ in $\alpha=0$ so that this exact value for $J$ can be inferred by developing
the second derivative for small values of $\alpha$, here this is no longer true, as we find that the first inflection point is obtained for
$J_3=0.5$, in correspondence of $\alpha=1.5$. This is related to the fact that the minimum of the free energy passes quite abruptly
(however continuously and it is ultimately due to a violation of the quadratic shape for the energetic term)
from values of $\theta$ not much larger than one half to larger values. We can then identify $J_3=0.5$ as the value
separating a weak cooperativity regime from a strong cooperativity one.
When $J_3$ is larger than this value, another zero in the second derivative appears, so that we have two inflection points.
Above the critical value $J_3=0.9$, instead, we have a unique inflection point below $\alpha = 1$.

For a critical interaction strength of $J_3 = 0.86 $, the system has a first order transition so that $ \theta$ is
continuous in $\alpha$ below a critical concentration $ \alpha^*(J_3) >  0 $ and it grows abruptly above this threshold.
Increasing $J_3$, this critical concentration becomes lower
up to negative values at certain $J_3$. In this case, due to the strong interaction, a concentration
smaller than $\alpha=1$ in sufficient to have this discontinuous transition between a negative value of $\theta -1/2$ and a positive one.
When $J_3$, and so the cooperativity is large, the critical line tends to be located at $\alpha =-\exp(3J_3/4)$.

\subsection{Case $p=4$}

As for p=2, in this case the energy is symmetric with respect to $\theta = 1/2$ when the concentration is equal to one,
so that also the free energy features this symmetry
\be
	F(\theta,\alpha) = \sup_{|\theta|}\Big(- 2J_4 (2 \theta - 1)^4 - \frac{1}{2} \log(\alpha)(2 \theta - 1) -s(\theta) \Big)
\ee
and thermodynamic stability requires its minimization (w.r.t. $\theta$) as
\be
 \theta = \frac{1}{2} + \frac{1}{2}\tanh(2J_4 (2\theta - 1)^3 + \frac{1}{2} \log(\alpha)).
\ee
When the interaction among sites is small, again, we have a binding fraction ruled by the concentration of free ligands and vanishing as the latter goes to zero.
For $J_4 = J_{c1} = 0.69 $ however, the system has a first order phase transition so that, coming from small concentrations,
there is a value $\alpha_c(J_4)<1$
for which $\theta$ changes abruptly to a larger value, which is smaller than $1/2$. 
Then $\theta(\alpha)$ varies continuously in the interval $\alpha_c(J_4) < \alpha < \alpha_c(J_4)'$,
passing through the point $(1, 1/2)$, and there is a new transition with a discontinuity
at an inverse value of $\alpha_c(J_4)$, $\alpha_c(J_4)' = \alpha_c(J_4)^{-1}$.
In fact for this range of interaction strengths $J_4$, and for concentrations around $\alpha_c(J_4)$,
the free energy has two local minima coming from the well matched competition
between the energetic and the entropic terms, the former preferring small, "ordered" $\theta$,
while the latter has always its maximum at $\theta=1/2$. The concentration tells us which is 
the global minimum and the two minima are equal at $\alpha_c$. The same happens
around the symmetrical value $\alpha_c(J_4)'$.  

\begin{figure}[htb]	
\includegraphics[width=10cm]{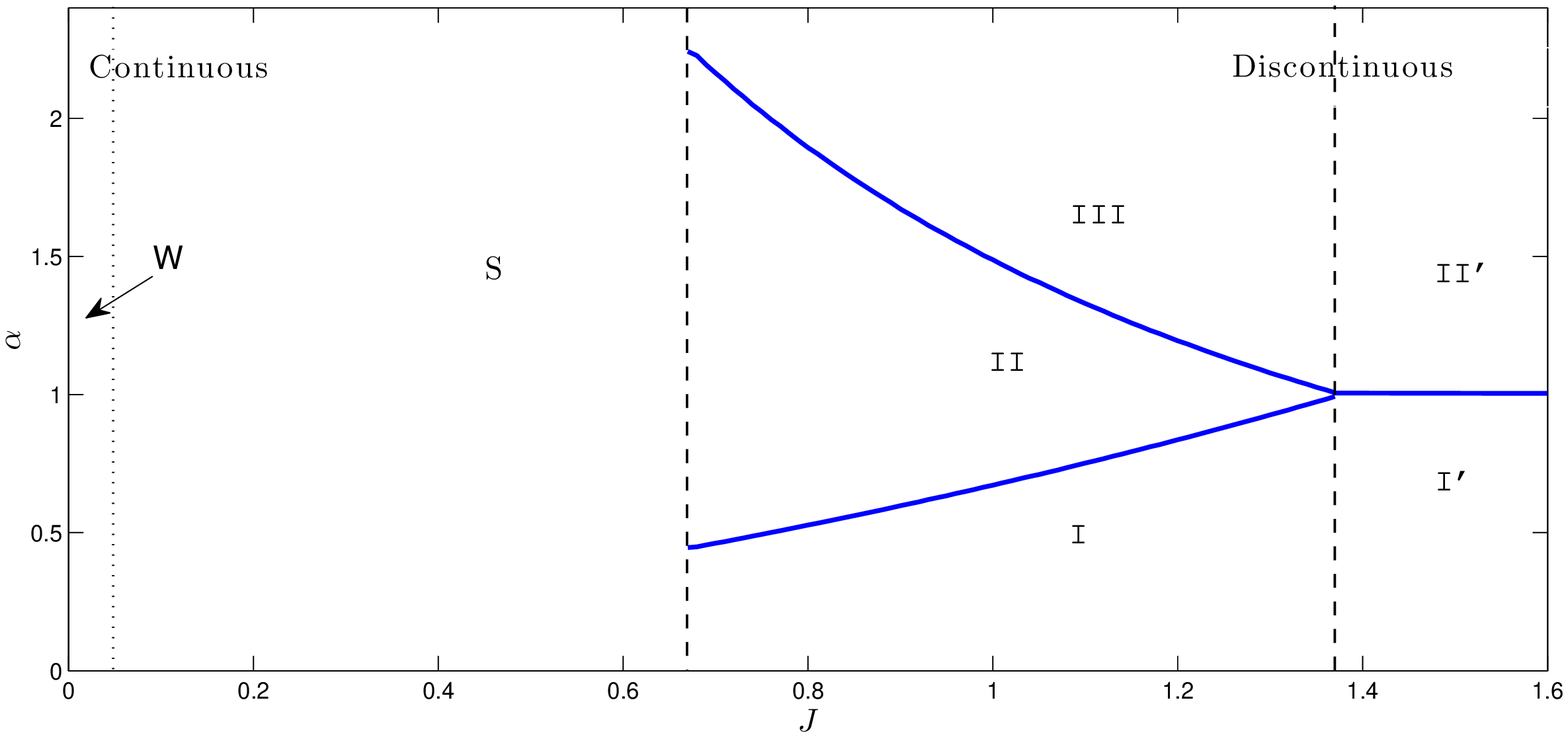} \\ \includegraphics[width=10cm]{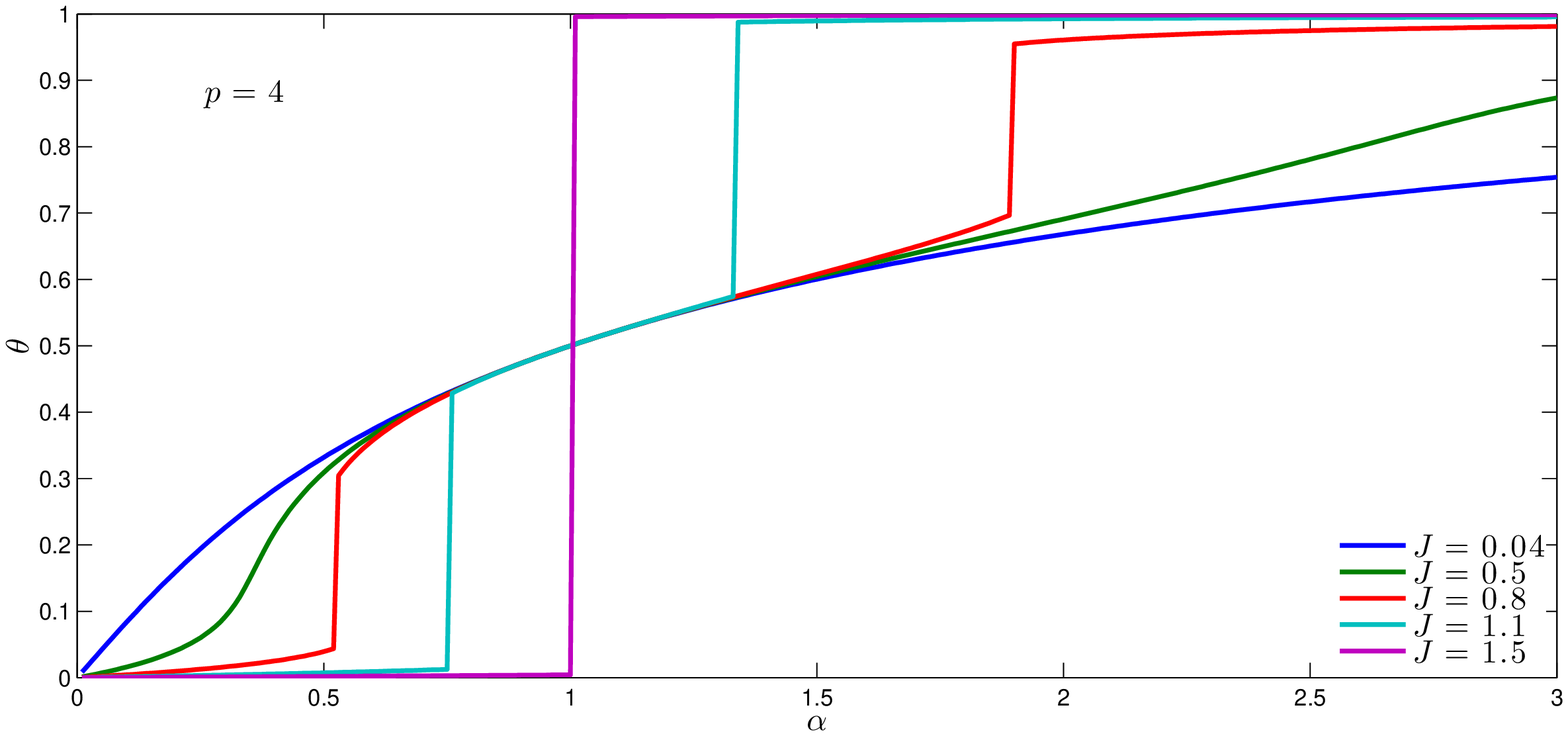}
\caption{Top: phases for the $p=4$ model.
Binding is continuous for $J_4<J_{c1}=0.69$, and it passes from a weak (W) to a strong (S) - both continuous - regime at $J_4=1/24$;
for $J_{c1}<J_4<J_{c2}=1.37$ isotherms have two discontinuities, one for a critical $\alpha_c<1$ and the other one for 
a $\alpha_c'=1/\alpha_c>1$(corresponding to a critical line, so
that we can identify three (I, II, III) continuous regions at a fixed $J_4$; when $J_4>J_{c2}$ however, there is only one large discontinuity at $\alpha=1$,
and the behavior is similar to the one observed in the $J_4>1$ phase for $p=2$(see Figure \ref{fig:fasip2}). 
Bottom: binding isotherms $\theta(\alpha)$ of a 4-bodies interacting system,
for different values of the interaction $J_4$.
The binding isotherm for $J_4=0.04$ (the blue line) has no inflections and represents a practically 
non cooperative system;
when $J_4=0.5$ (green)  the curve has an inflection point for a small concentration; 
for $J_4>J_{c1}$ two inflection points appear and the isotherms develop two discontinuities corresponding respectively
to the concentrations $\alpha_1$ and $\alpha_2$ such that $\alpha_1 \alpha_2 = 1$, see the
curves for $J_4=0.8$(red) and $J_4=1.1$(cyan);
for interactions larger than $J_{c2}$, the system has just one discontinuity for $\alpha = 1$ 
and the binding isotherm tends to a step function, as for the purple line corresponding to $J_4=1.5$.
 \label{fig:p4}}
 \end{figure}

The critical concentration $\alpha_c$ tends to one with the interaction strength growing,
and for $J_4 = J_{c2}=1.37$ the system becomes ferromagnetic.
In this case the mutual interaction is the overwhelming force driving the system,
and, as in the previous $p=2$ case, the binding isotherm has a discontinuity when passing
from $\alpha<1$ to $\alpha>1$, changing abruptly from a large negative value (depending on $J_4$)
to a positive, opposite value (see Figure \ref{fig:p4}).
It is interesting to note that the values of $\alpha_c$ and $\alpha_c'$ depend on $J_4$ and even for $p=3$ we find a critical concentration
depending on $J_4$: because this dependence is absent in the $p=2$ case, an experimental $J_4$-dependent critical concentration
may indicate that a multiple interaction effect is acting (this can be seen for example in \cite{murase}).

As for $p=2,3$,  we can determine exactly the value $J_4=1/24$ as the one for which
a first inflection point appears,
so that the binding isotherm passes from a weak to a strong cooperativity region. In fact, as for the $p=2$ case (see eq. \ref{ddf0})
by gaining the second derivative of the binding isotherm for small $\alpha$ one can see that the leading term is proportional to $(1-24J_4)$.
The isotherm has a unique inflection point until $J_4 = 0.45$, then a second point appears at $\alpha=2.4$, which for larger $J_4$ splits in two points,
so that for $0.45<J_4<J_{c1}$ we have three inflections. The ones corresponding to the smaller and to the larger values of $\alpha$ disappear
when $J_{c1}$ is reached and, instead of these inflections, the isotherm has two discontinuities.

A discontinuous behavior which could be explained in term of two and multisite interactions between hosting sites
has been observed in the binding isotherm of long chain alkyl sulphates and sulphonates to the protein bovin serum albumin,
and in the adsorption isotherm of alkylammonium chlorides chains on the biotite surface \cite{langmuir}.
Typically, if there are multi-site interactions (with $p >2$) it is very likely that also two-bodies interactions are present,
and one should consider the several possible interactions with different strengths $J_p$ in the energy term,
while for the sake of clearness we considered them separately.

Before concluding some remarks are in order:
\newline
As we have shown, the phenomenology pertaining to systems with $p$-body interactions depends sensitively on $p$ and this allows to infer information about the properties of the system under study, starting from its reaction rate.

On the other hand, we also notice that when looking at the most likely configuration, that is the most likely value of $\theta$ for a given parameter set $(p, J, \alpha)$, we find that it exists and it is unique: if we, for the sake of simplicity, focus just on even $p>2$ values, we can see that the amount of solutions is constant, namely, independent by the amount and complexity of the cooperating/anti-cooperating binding sites. More precisely, the self-consistent relation in Eq.~$28$ may allow for one or two distinct solutions, but only one is a global minimum for the free energy: the minima of the free energy do not scale, but simply shift, with $p$, while the global minimum is always unique.
This argument may be applied to the problem concerning the folding of (long) proteins, whose secondary and tertiary structure is essentially always the same, despite a large number of multi-attachment is in principle possible \cite{parisipagnani}.

\section{Conclusions and perspectives}

In this paper we pursued the strand early paved by J. J. Thompson \cite{thompson} in modeling chemical kinetics with statistical mechanics techniques: highlighting the emergent properties in this field, namely cooperation among binding sites,
we relaxed the original assumption of dealing with geometric  models (i.e. linear chains)  in favor of mean field models as the latter may properly account also  other  experimental findings (as discontinuous jumps or folding) and benefit of a broad tunable scenarios able to recover essentially all the definition of cooperativity introduced in literature.
 
After a streamlined introduction to the modern definitions of "cooperativity" in chemical kinetics, we developed a one-to-one mapping between chemical variables (binding sites, substrate, etc) and the statistical mechanics counterparts. Then, we applied statistical mechanics prescriptions, that is the variational principle (minimization of the free energy) and we found an expression for the reaction rate $\theta$ (representing the number of occupied sites) as a function of the substrate concentration $\alpha$. In doing so the interaction constant $J$ between binding sites is treated as a free (microscopic) parameter. We show that by tuning $J$ we are able to recover a plethora of different behaviors, ranging from non-cooperative ($J=0$), weak cooperativity ($J<1/4$) and strong cooperativity ($J>1/4$). Our results are also shown to be consistent with the number of definitions of cooperativity (e.g. Hill, Koshland, or range), previously introduced.
Furthermore, we found a universal criterion to distinguish between weak and strong cooperativity, that is, by looking at the inflections of the second derivative of the free energy. Indeed, we found that the existence of cooperativity (i.e. is a positive interaction constant) is not sufficient for the existence of a flex in the reaction rate. Otherwise stated, a non-sigmoidal law does not necessarily mean that no cooperativitiy is at work. Such situations are here referred to as cooperativity of weak degree.

There are several directions along which this work can be extended: for example including a well defined topology for binding sites, which would account to more realistic interaction scalings among the binding sites. Also, non-strictly-positive couplings would realize a spin-glass system which, as well known from statistical mechanics,
significantly enriches the phenomenology, to be possibly mapped in terms of reaction kinetics.

\section*{Acknowledgements}
The research belongs to the strategy of exploration funded by the
FIRB project RBFR08EKEV which is acknowledged.

\bibliography{mfcoop}
\bibliographystyle{plain}

\end{document}